# Modélisation des déformations de maturation de la fibre

**François Sassus*— Tancrède Alméras*,** — Joseph Gril* — Hiroyuki Yamamoto****

** Laboratoire de Mécanique et Génie Civil, Université Montpellier II*
*CC-048 Place E. Bataillon, 34095 Montpellier Cedex 5*
*jgril@lmgc.univ-montp2.fr*
*** School of Bio-agricultural Science, Nagoya University, 464-8601 Nagoya, Japan*

RESUME. *Le mécanisme de genèse des déformations de maturation du bois est étudié au moyen d'une analyse élastique de la paroi cellulaire considérée comme un tube composite, dont les couches sont constituées d'une matière ligno-cellulosique variant dans sa composition et son angle microfibrillaire. Pour chaque couche, la cinétique de déformation induite est supposée synchronisée avec celle de la rigidification. Dans ce cas des variations de la composition concomitantes avec celles de l'angle doivent être invoquées pour expliquer les variations observées de la déformation de maturation.*

ABSTRACT. *The generation mechanism of growth strains is investigated theoretically by developing an elastic analysis of a wood cell viewed as a composite tube, with layers composed of a ligno-cellulosic material of varying composition and microfibrillar orientation. For each layer, the kinetics of induced strain is assumed synchronous with that of rigidification. In this case, variations of polymer behaviour concomitant with orientation must be invoked to account for observed variations of the growth strain.*

## 1. Introduction

On peut se demander dans quelle mesure un modèle de fibre peut rendre compte des propriétés macroscopiques du bois. La fibre des feuillus (ou la trachéide pour les résineux) est le constituant principal du bois. Certaines propriétés ont un caractère additif, par exemple la masse volumique du bois va dépendre de la masse volumique de chaque constituant (fibre, vaisseau, rayon, …) et de leur proportion. Ceci n'est pas le cas des modules d'élasticité par exemple, qui eux vont aussi dépendre de l'arrangement des constituants [CAV 68]. La rigidité des vaisseaux étant très inférieure à celles des fibres (ils sont constitués surtout de vide), le comportement dans la direction correspondant à la mise en parallèle des deux est quasiment celui des fibres. Cette direction est ici la direction longitudinale du bois, pour laquelle dans notre raisonnement le retrait des fibres rend compte directement du retrait du bois macroscopique. Resterait alors à analyser le comportement des rayons. On peut avancer qu'ils sont peu importants en proportion. La direction transverse doit tenir compte de la disposition des parois des différents éléments qui décrit des arrangements allant de la situation en parallèle à la situation en série. Cette analyse est complexe et permet de prédire une anisotropie transverse du bois avec comme hypothèse que les constituants des parois sont isotropes dans ce plan [WAT 96]. C'est pourquoi la suite de notre exposé se limitera à la direction longitudinale.

D'un point de vue historique, les déformations de la fibre ont d'abord été abordées par des considérations géométriques. Elles sont basées sur une schématisation de la géométrie de la fibre. Seule la couche S2 est généralement prise en compte dans ce type de modèle. Par la suite, les modèles de déformations de la fibre ont été abordés par une description mécanique. Ils sont basés sur la description de la matière ligneuse. Barber et Meylan [BAR 64] développent un modèle de "matrice renforcée", la matrice représentant la partie amorphe de la matière (lignines, …) dans laquelle des microfibrilles de cellulose jouent un rôle de renfort car très rigide. La description du comportement mécanique des deux constituants et de leur agencement permet de décrire les déformations de séchage [CAV 72]. Ces modèles ont été développés pour décrire de manière plus complète la géométrie et l'hétérogénéité de la cellule de bois [YAM 95, YAM 98].

Les modèles mécaniques développés dans la littérature mettent en évidence une forte relation entre la structure de la matière ligneuse et les déformations de la fibre. Ces résultats sont confirmés par les données expérimentales. La description développée dans ce paragraphe, est inspirée d'un modèle antérieur [YAM 95]. Ce modèle a pour point de départ la description de la matière ligneuse. Elle est décrite comme le renforcement d'une matière amorphe constituée principalement de lignines par un réseau de microfibrilles de cellulose cristalline.

## 2. Modèle mécanique de la matière ligneuse

### *2.1 Structure*

La matière ligneuse des couches de la paroi cellulaire est constituée de deux matériaux de base : la matrice et le réseau microfibrillaire. La matrice regroupe l'ensemble des constituants macromoléculaires amorphes du bois, principalement les lignines. Elle inclut la cellulose non cristalline et les hémicelluloses, même si leur état peut être considéré comme organisé avec une direction préférentielle le long des microfibrilles de cellulose ainsi que les autres constituants minoritaires (extractibles). Le réseau microfibrillaire représente les microfibrilles de cellulose cristalline “pontées” par des liaisons hydrogènes. Les deux composants sont “entrelacés” (figure 1). En mélange, il n'y a pas d'espace à l'échelle macroscopique où un seul des deux composants est présent, ce qui est le cas dans un matériau stratifié par exemple. La matrice constitue un ciment qui occupe l'espace entre les microfibrilles.

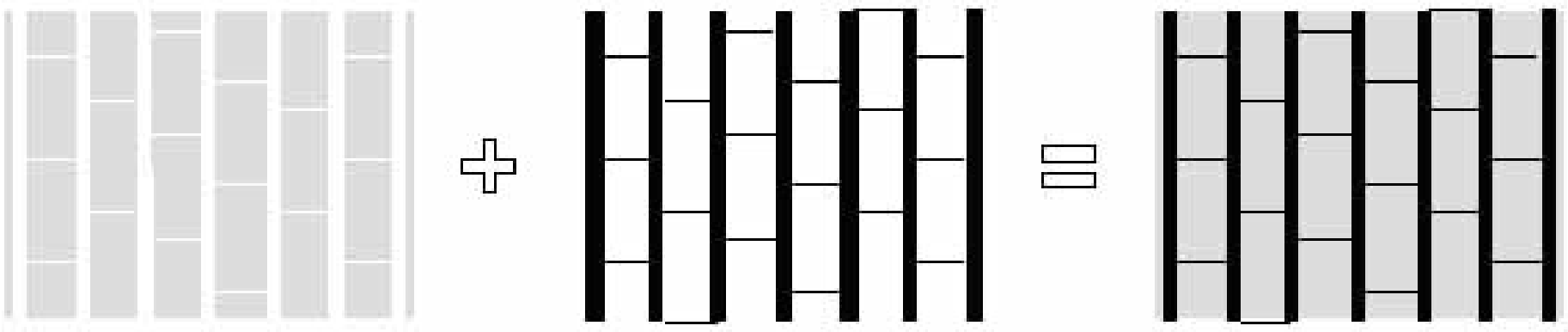

**matrice** **réseau microfibrillaire** **matière ligneuse**

**Figure 1.** *description de la couche comme l'entrelacement de la matrice et du réseau microfibrillaire*

#### *2.1.1. Comportement mécanique des constituants*

Le retrait et la maturation induisent des déformations dans les deux constituants des parois. La matrice et le réseau suivent des lois de comportement élastique avec déformations induites de la forme : $\underline{\sigma}=\underline{C}(\underline{\varepsilon}-\underline{\alpha})$, où $\underline{\sigma}$ désigne le tenseur des contraintes, $\underline{\varepsilon}$ le tenseur des déformations et C le tenseur de rigidité élastique. Le tenseur $\underline{\alpha}$ est analogue à une déformation, il est appelé tenseur des **déformations potentielles**. Lorsque la déformation exprimée $\underline{\varepsilon}$ est égale à la déformation potentielle $\underline{\alpha}$, le processus ne génère aucune contrainte dans le matériau. L'exposant f désignera dans la suite le réseau microfibrillaire alors que m désignera la matrice.

La matrice est considérée comme un matériau isotrope : sa rigidité est définie par le module d'Young $E_m$, le coefficient de Poisson $\nu_m$ et le module de cisaillement $G^m=E^m/[2(1+\nu^m)]$ ; sa déformation potentielle s'écrit comme un tenseur de déformation isotrope $\underline{\alpha}^m = \alpha^m\ \underline{Id}$. Le réseau microfibrillaire est décrit comme un matériau isotrope transverse (figure 2). Ses caractéristiques mécaniques sont données par : la direction l des microfibrilles qui permet de définir la direction principale et le plan isotrope (t), $E^f_l$ le module d'élasticité dans la direction de la microfibrille, $E^f_t$ le module d'élasticité dans le plan orthogonal aux microfibrilles, les coefficients de Poisson dans le plan tl $\nu^f_{tl}$ et dans la plan transverse $\nu^f_t$ ainsi que les modules de cisaillement $G^f_{tl}$ et $G^f_t$.

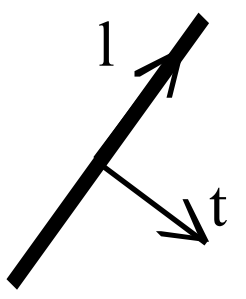

**Figure 2.** *Repère local de la microfibrille*

La déformation potentielle est définie comme une déformation isotrope transverse $\alpha^f_{tt}$ et $\alpha^f_{ll}$ dans les directions principales, pour laquelle on suppose les déformations de cisaillement nulles. Les tenseurs de rigidité et de déformation sont exprimés dans le repère isotrope transverse en notations vectorielles :

$$(\underline{\underline{C}}^m)^{-1} = \begin{bmatrix} \frac{1}{E^m} & -\frac{\nu^m}{E^m} & -\frac{\nu^m}{E^m} & & & \\ -\frac{\nu^m}{E^m} & \frac{1}{E^m} & -\frac{\nu^m}{E^m} & & & \\ -\frac{\nu^m}{E^m} & -\frac{\nu^m}{E^m} & \frac{1}{E^m} & & & \\ & & & \frac{1}{G^m} & & \\ & & & & \frac{1}{G^m} & \\ & & & & & \frac{1}{G^m} \end{bmatrix} \quad \text{et} \quad \underline{\alpha}^m = \begin{bmatrix} \alpha^m \\ \alpha^m \\ \alpha^m \\ 0 \\ 0 \\ 0 \end{bmatrix}$$

$$\left(\underline{\underline{C}}^f\right)^{-1} = \begin{bmatrix} \frac{1}{E_t^f} & -\frac{\nu_t^f}{E_t^f} & -\frac{\nu_{lt}^f}{E_l^f} & & & \\ -\frac{\nu_t^f}{E_t^f} & \frac{1}{E_t^f} & -\frac{\nu_{lt}^f}{E_l^f} & & & \\ -\frac{\nu_{lt}^f}{E_l^f} & -\frac{\nu_{lt}^f}{E_l^f} & \frac{1}{E_l^f} & & & \\ & & & \frac{1}{G_t^f} & & \\ & & & & \frac{1}{G_{tl}^f} & \\ & & & & & \frac{1}{G_{tl}^f} \end{bmatrix} \quad \text{et} \quad \underline{\alpha}^f = \begin{bmatrix} \alpha_t^f \\ \alpha_t^f \\ \alpha_l^f \\ 0 \\ 0 \\ 0 \end{bmatrix}$$

Le pourcentage de vide dans la matrice est noté x. C'est aussi le pourcentage de matière dans le réseau microfibrillaire. Les rigidités des matériaux peuvent être estimées à partir du pourcentage et des caractéristiques des macromolécules qui les constituent. Ainsi le module d'Young et de cisaillement de la matrice est calculé à partir de ceux de la lignine $E^m_0$ et $G^m_0$ et de sa proportion volumique 1-x.

$$E^m = (1-x)\, E^m_0 \qquad G^m = (1-x)\, G^m_0$$

Pour le réseau microfibrillaire, les rigidités sont de natures différentes dans la direction principale et le plan transverse. L'estimation du module d'Young longitudinal peut être faite à partir du module de la cellulose cristalline $E^f_c$ et de la proportion x.

$$E^f_l = x\, E^f_c$$

Les autres composantes des rigidités du réseau microfibrillaire restent difficiles à évaluer. Le module d'Young transverse est caractéristique des liaisons transverses entre les microfibrilles de cellulose. Ces liaisons sont moins nombreuses que les liaisons chimiques des lignines ; pour cette raison, le module transverse du réseau microfibrillaire est supposé faible devant celui de la lignine :

$$E_t^f \ll E^m$$

Les modules de cisaillement des deux constituants peuvent par contre être du même ordre.

*2.1.2. Superposition des constituants*

La matrice et le réseau microfibrillaire sont représentables par deux matériaux poreux superposés : le vide dans le réseau microfibrillaire est occupé par la matrice et vice versa (Figure 3). On suppose que la superposition des deux matériaux est telle que leur déformation est la même et égale à la déformation de la couche et que la contrainte de la couche est la somme des contraintes de la matrice et du réseau microfibrillaire :

$$\varepsilon = \varepsilon^m = \varepsilon^f \qquad \text{et} \qquad \sigma = \sigma^f + \sigma^m$$

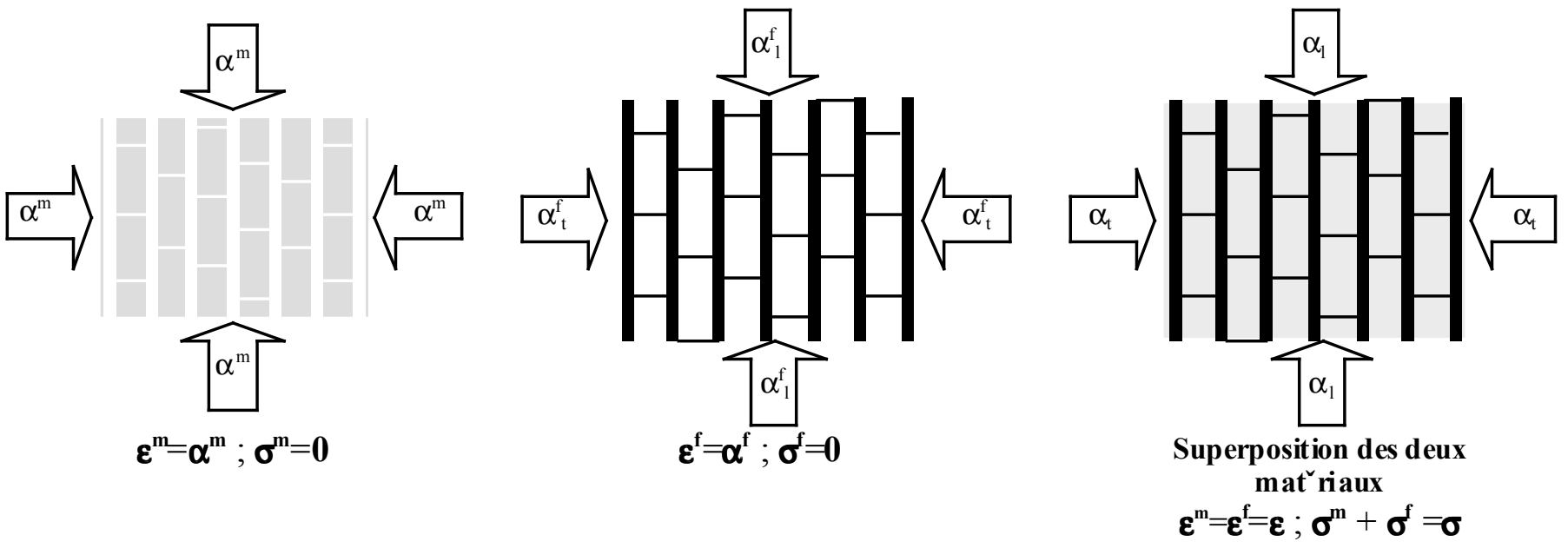

**Figure 3.** *Schématisation du principe de superposition des deux constituants de la couche.*

Comparativement aux images classiques de matériaux multicouches, cette hypothèse représente l'assemblage de deux composants en parallèle dans toutes les directions. Le principe de superposition des déformations et des contraintes permet d'écrire comme loi de comportement pour la couche :

$$\underline{\sigma} = \underline{\underline{C}}^f(\underline{\varepsilon}-\underline{\alpha}^f) + \underline{\underline{C}}^m(\underline{\varepsilon}-\underline{\alpha}^m) = \underline{\underline{C}}(\underline{\varepsilon}-\underline{\alpha})$$

C'est aussi une loi de comportement élastique avec déformations induites. La rigidité et la déformation potentielle sont :

$$\underline{\underline{C}} = \underline{\underline{C}}^f + \underline{\underline{C}}^m \text{ et } \underline{\alpha}=\underline{\underline{C}}^{-1}[\underline{\underline{C}}^f\underline{\alpha}^f + \underline{\underline{C}}^m\underline{\alpha}^m] \qquad [1]$$

La matière ligneuse ainsi construite par superposition est isotrope transverse dans le repère lié au réseau microfibrillaire. Les caractéristiques de la loi de comportement élastique avec déformations induites s'expriment dans ce repère sous forme vectorielle :

$$\underline{\underline{C}} = \begin{bmatrix} C_t & C_{tt} & C_{tl} & & & \\ C_{tt} & C_t & C_{tl} & & & \\ C_{tl} & C_{tl} & C_l & & & \\ & & & C_{\tau\lambda} & & \\ & & & & C_{\tau\lambda} & \\ & & & & & C_{\lambda\lambda} \end{bmatrix} \text{ et } \underline{\alpha} = \begin{bmatrix} \alpha_t \\ \alpha_t \\ \alpha_l \\ 0 \\ 0 \\ 0 \end{bmatrix}$$

*2.1.3. Comparaison avec les modèles "série parallèle"*

Les constantes élastiques construites par superposition peuvent être comparées avec des modèles plus classiques et naturels de description de la matière ligneuse comme l'empilement des constituants. Un modèle plus récent [HAR 98] détermine les caractéristiques de la couche pariétale par homogénéisation des caractéristiques des constituants : (1) Cellulose cristalline, (2) polyose associé à la cellulose, (3) polyose associé à la lignine et (4) lignine. Les deux premiers sont considérés comme des matériaux isotropes transverses, les deux derniers comme isotropes. Les résultats obtenus sont confirmés par un calcul par éléments finis et utilisent comme composition et rigidités des constituants (Tableau 1).

**Tableau 1.** *Données décrivant la composition de la paroi d'une trachéide de résineux.*

| | | (1) | (2) | (3) | (4) |
|---|---|---|---|---|---|
| Proportion volumique | dans M+P | 0,11 | 0,08 | 0,06 | 0,75 |
| | dans $S_1$ | 0,37 | 0,30 | 0,05 | 0,28 |
| | dans $S_2$ | 0,45 | 0,33 | 0,01 | 0,21 |
| | dans $S_3$ | 0,34 | 0,36 | 0,00 | 0,30 |
| Rigidités (GPa) | $E_t$ | 17,7 | 6,81 | 3,18 | 4,18 |
| | $E_l$ | 128 | 15,9 | - | - |
| | $\nu_{lt}$ | 0,33 | 0,33 | - | - |
| | $G_{lt}$ | 6,00 | 2,38 | 1,22 | 1,61 |
| | $G_{tt}$ | 6,00 | 2,38 | - | - |

Les rigidités obtenues par le modèle "parallèle-série" sont comparées à celles de notre modèle de superposition (Tableau 2). La proportion de matrice est prise comme la somme de (2) (3) et (4).

Les modèles "série-parallèle" donnent des valeurs distinctes pour la direction radiale et tangentielle de la paroi. Elles sont peu différentes et la moyenne est identifiée aux directions de notre modèle. Hormis l'écart du coefficient de Poisson $\nu_{tl}$, les données numériques sont proches. Ceci semble indiquer que l'agencement parallèle qui est le seul considéré dans notre modèle de superposition est le principal responsable du comportement mécanique de la matière ligneuse.

***Tableau 2.*** Constantes élastiques calculées par [a] modèle "série-parallèle" ;[b] modèle de superposition

| | $E_t$ | | $E_l$ | | $\nu_{tt}$ | | $\nu_{tl}$ | | $G_{tt}$ | | $G_{tl}$ | |
|---|---|---|---|---|---|---|---|---|---|---|---|---|
| | [a] | [b] | [a] | [b] | [a] | [b] | [a] | [b] | [a] | [b] | [a] | [b] |
| S3 | 8,21 | 8,43 | 50,36 | 46,29 | 0,39 | 0,46 | 0,33 | 0,32 | 2,68 | 3,08 | 2,83 | 3,10 |
| S2 | 9,51 | 10,44 | 63,96 | 59,91 | 0,39 | 0,47 | 0,33 | 0,32 | 2,96 | 3,56 | 3,20 | 3,58 |
| S1 | 8,28 | 9,38 | 53,10 | 50,00 | 0,38 | 0,46 | 0,33 | 0,32 | 2,66 | 3,21 | 2,84 | 3,23 |
| M+P | 5,10 | 5,88 | 18,43 | 17,80 | 0,38 | 0,41 | 0,31 | 0,31 | 1,88 | 2,08 | 1,95 | 2,09 |

Cette conclusion est valable pour le cas considéré ici, qui est la matière ligneuse à l'état mature. Mais elle ne doit pas masquer une différence fondamentale entre les constituants et le réseau formé par un constituant. Le modèle série parallèle décrit les constituants chimiques du bois et leur agencement, alors que le modèle de superposition décrit les réseaux que forment les constituants. Cette différence peut être illustrée par les caractéristiques du bois avant maturation, c'est à dire avant lignification. La contribution de la lignine peut alors être décrite par un matériau de rigidité quasi nulle. Dans ces conditions, le modèle de superposition attribue à la paroi la rigidité et les déformations potentielles du réseau microfibrillaire ($\underline{\underline{C}}^m=0$ dans l'équation 1). Notamment la rigidité transverse du réseau microfibrillaire se confond alors avec celle de la paroi ; si nous utilisions les valeurs de la microfibrille (Tableau 1), cela aboutirait à confondre le comportement de la microfibrille avec celui de la paroi. Pour pallier cet inconvénient, nous avons choisi un jeu de paramètres qui permet de rendre compte des caractéristiques mécaniques du bois macroscopique et qui assure néanmoins une rigidité transverse du réseau microfibrillaire petite devant celle de la matrice (Tableau 3).

**Tableau 3.** *Caractéristiques mécaniques des constituants et de la couche S2 déduites par le modèle de superposition*

| | Matrice | Réseau microfibrillaire | Couche S2 |
|---|---|---|---|
| $E_t$ | 10 | 0.5 | 6.20 |
| $E_l$ | - | 100 | 55.51 |
| $\nu_{lt}$ | - | 0.2 | 0.30 |
| $G_{lt}$ | - | 5 | 4.37 |
| $G_{tt}$ | 3.85 | 0.03 | 2.13 |

Ces nouveaux paramètres correspondent à une diminution du module d'élasticité longitudinal du réseau microfibrillaire et une augmentation pour la matrice. La modification principale est la forte diminution du module transverse du réseau microfibrillaire, compensée par une surévaluation du module de la matrice. Les caractéristiques mécaniques déduites par le modèle de superposition pour la couche S2 sont proches de celles évoquées précédemment.

Cet artefact est liée au caractère isotrope imposé à la matrice et aurait pu être évité en supposant une rigidité de la matrice plus élevée dans la direction t que dans la direction l. En réalité, c'est la matière constitutive de la matrice qui pourrait être considérée comme isotrope, et non pas la contribution du réseau matriciel.

### *2.2 Structure multicouche de la fibre de bois*

#### *2.2.1. Géométrie de la fibre de bois*

La fibre est représentée par une succession de couches cylindriques coaxiales. La lamelle moyenne (M) et la paroi primaire (P) sont de faibles épaisseurs et sont majoritairement constituées de lignines ; dans notre modèle elles sont regroupées pour former la couche externe M+P (figure 4). Viennent ensuite les parois secondaires $S_1$ et $S_2$ plus riches en cellulose. A ces trois couches, une quatrième peut être ajoutée à l'intérieur pour une couche gélatineuse dans le cas d'une fibre G de bois de tension. La couche $S_3$ n'est pas représentée dans le modèle car de faible épaisseur. La géométrie transverse de la fibre est paramétrée par les rayons ($r_i$) pour i=0 à n ; $r_0$ est le rayon externe de la fibre, $r_1$ le rayon interne de M+P etc…

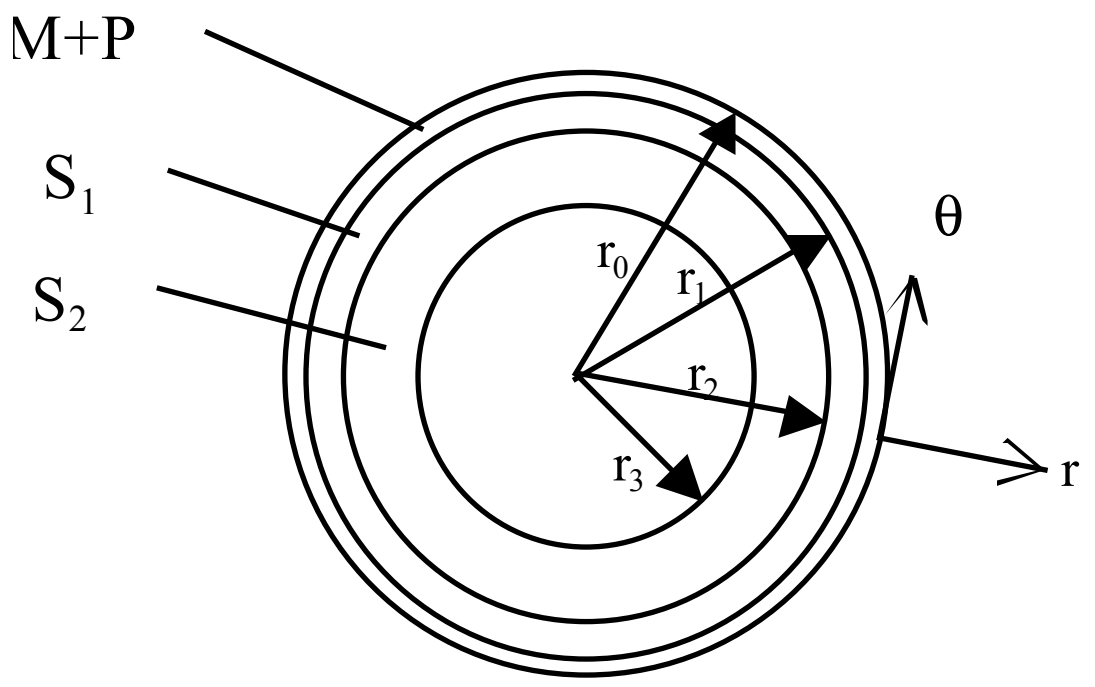

**Figure 4.** *Géométrie de la fibre à trois couches dans le plan transverse*

*2.2.2. Structure des parois*

Les microfibrilles de cellulose ont une direction privilégiée dans les couches $S_1$, $S_2$ et $S_3$. Elles forment des hélices parallèles entre elles. L'angle de la tangente de l'hélice avec l'axe de la fibre est appelé angle des microfibrilles (AMF) dans la couche. {Les microfibrilles peuvent être croisées, comme dans la couche $S_1$ où schématiquement la moitié des microfibrilles font un angle $+\varphi$ avec l'axe de la fibre alors que l'autre moitié fait un angle opposé $-\varphi$.} L'angle des microfibrilles est caractéristique de la couche et de la nature du bois considéré. Le Tableau 4 rapporte les données utilisées par Schniewind [SCH 72] pour les angles des microfibrilles ($\varphi$) et les épaisseurs relatives (P) de chaque couche pour les trachéides de résineux et les données communes pour les fibres.

**Tableau 4:** *Angle des microfibrilles pour chaque couche en fonction de la nature du bois*

| | M+P | | $S_1$ | | $S_2$ | | $S_3$ | |
|---|---|---|---|---|---|---|---|---|
| | P | $\varphi$ | P | $\varphi$ | P | $\varphi$ | P | $\varphi$ |
| Trachéide de bois initial | 0.1 | 90° | 0.3 | 80° | 0.2 | 40° | 0.4 | 70° |
| Trachéide de bois final | 0.03 | 90° | 0.07 | 80° | 0.8 | 10° | 0.1 | 70° |

Seul l'angle des microfibrilles dans la couche $S_2$ varie notablement en fonction de la nature du bois. AMF désignera dans la suite l'angle des microfibrilles dans cette couche.

Deux types de cellules sont modélisées dans la suite (Figure 5) :

- fibre de bois normal constituée de (M+P), $S_1$ et $S_2$. Ce modèle représente aussi la trachéide de bois de compression des résineux ou la fibre de bois de tension des espèces ne formant pas de couche gélatineuse dans leur bois de tension.
- fibre gélatineuse constituée de (M+P), $S_1$, $S_2$ et G

*2.2.3. Comportement des parois dans le repère de la fibre*

La couche est décrite par la superposition de la matrice isotrope et du réseau microfibrillaire isotrope transverse dans {tl}. Son comportement est donc isotrope transverse dans ce repère. Dans le repère cylindrique local de la fibre, si les microfibrilles font une angle différent de 0 et 90°, alors ce repère local ne définit plus trois plans de symétrie pour le matériau, la loi de comportement de la couche considérée n'est plus orthotrope cylindrique mais anisotrope. De même, pour un tel angle de microfibrille, la déformation potentielle en cisaillement dans le plan $\theta z$ de la couche est non nulle.

Les caractéristiques de la couche sont d'abord exprimées dans le repère isotrope transverse défini par le réseau microfibrillaire, puis dans le repère de la fibre par rotation. Les rigidités font apparaître un couplage entre les déformations normales et les efforts de cisaillement dans le plan θz alors que les déformations potentielles comprennent un terme de cisaillement $\delta_{\theta z}$ dans le plan θz..

$$\underline{\underline{C}} = \begin{bmatrix} C_{rr} & C_{r\theta} & C_{rz} & C_{r\rho} & & \\ C_{\theta r} & C_{\theta\theta} & C_{\theta z} & C_{\theta\rho} & & \\ C_{zr} & C_{z\theta} & C_{zz} & C_{z\rho} & & \\ C_{\rho r} & C_{\rho\theta} & C_{\rho z} & C_{\rho\rho} & & \\ & & & & C_{\tau\tau} & C_{\tau\lambda} \\ & & & & C_{\tau\lambda} & C_{\lambda\lambda} \end{bmatrix}_{r,\theta,z} \qquad \underline{\alpha} = \begin{bmatrix} \alpha_r \\ \alpha_\theta \\ \alpha_z \\ \delta_{\theta z} \\ 0 \\ 0 \end{bmatrix}$$

où r, θ, z désignent les directions principales du repère cylindrique, et ρ, τ, λ les plans de cisaillement θz, rz, et rθ, respectivement.

La rotation des caractéristiques de la matière ligneuse exprimées dans le repère isotrope transverse est exprimée sous forme matricielle (équation 2). On désigne par s le sinus de l'AMF et par c son cosinus :

$$\begin{bmatrix} \alpha_r \\ \alpha_\theta \\ \alpha_z \\ \delta_{\theta z} \\ \delta_{zr} \\ \delta_{r\theta} \end{bmatrix} = \begin{bmatrix} 1 & & & & & \\ & c^2 & s^2 & & & \\ & s^2 & c^2 & & & \\ & -2sc & 2sc & & & \\ & & & & & \\ & & & & & \end{bmatrix} \begin{bmatrix} \alpha_t \\ \alpha_t \\ \alpha_l \\ 0 \\ 0 \\ 0 \end{bmatrix}$$

$$\begin{bmatrix} C_{rr} \\ C_{\theta\theta} \\ C_{zz} \\ C_{\theta z} \\ C_{zr} \\ C_{r\theta} \\ C_{r\rho} \\ C_{\theta\rho} \\ C_{z\rho} \\ C_{\rho\rho} \\ C_{\tau\tau} \\ C_{\lambda\lambda} \\ C_{\tau\lambda} \end{bmatrix} = \begin{bmatrix} 1 & & & & & & & & \\ & c^4 & s^4 & 2c^2s^2 & & & 4c^2s^2 & & \\ & s^4 & c^4 & 2c^2s^2 & & & 4c^2s^2 & & \\ & c^2s^2 & c^2s^2 & c^4+s^4 & & & -4c^2s^2 & & \\ & & & & c^2 & s^2 & & & \\ & & & & s^2 & c^2 & & & \\ & & & & cs & -cs & & & \\ & -c^3s & cs^3 & cs\left(c^2-s^2\right) & & & 2cs\left(c^2-s^2\right) & & \\ & -cs^3 & c^3s & cs\left(s^2-c^2\right) & & & 2cs\left(s^2-c^2\right) & & \\ & c^2s^2 & c^2s^2 & -2c^2s^2 & & & c^2-s^2 & & \\ & & & & & & & c^2 & s^2 \\ & & & & & & & s^2 & c^2 \\ & & & & & & & sc & -sc \end{bmatrix} \begin{bmatrix} C_t \\ C_t \\ C_l \\ C_{tl} \\ C_{tl} \\ C_{tt} \\ C_{\tau'\lambda'} \\ C_{\tau'\lambda'} \\ C_{\lambda'\lambda'} \end{bmatrix} \quad [2]$$

où τ' et l' représentent les plans de cisaillement tl et tt, respectivement.

### *2.3 Modèle de fibre unique*

#### *2.3.1. Déformations planes généralisées et axisymétrie*

Le rapport hauteur sur diamètre de la fibre est de l'ordre de 100. Cet élancement permet de la considérer comme infinie le long de son axe. Suffisamment loin des extrémités, la résolution peut être faite dans le cadre des déformations planes généralisées. L'axisymétrie permet d'écrire les champs de déformations comme fonction de r uniquement et assure la nullité du déplacement suivant θ. La déformation dans la direction z est constante et notée $\Delta\varepsilon_z$. De plus la symétrie de révolution par rapport à l'axe de la fibre ainsi que la symétrie du chargement implique une solution indépendante de la position angulaire θ. Les champs solutions ne dépendent que de la position radiale r.

On notera dans le repère cylindrique {r,θ,z} et pour chaque couche i de la fibre :

$\underline{u}^i = (u^i_r, u^i_\theta=0, u^i_z)$ la solution en déplacement,

$\underline{\sigma}^i$ la solution en contrainte en notation vectorielle.

#### *2.3.2. Restriction de cisaillement*

Considérons tout d'abord le cas d'une **cellule isolée**. La déformation potentielle de la couche dans le cas d'un angle de microfibrille dans $S_2$ différent de 0 et ±90° comporte un terme de cisaillement. La résolution analytique du problème de déformation de la fibre isolée serait possible. Par exemple, Archer [ARC 86] propose une solution de déformations induites dans un matériau orthotrope dont la direction principale fait un angle φ avec l'axe de la tige, appliqué au cas des déformations de maturation dans le cas d'un angle du fil. Les déformations ainsi trouvées mettent en évidence un cisaillement dans le plan θz constant et proportionnel à sin(2φ) donc nul uniquement pour un angle du fil de 0 ou 90°. Cette déformation de cisaillement $\gamma_{\theta z}$ correspond à une torsion de la tige.

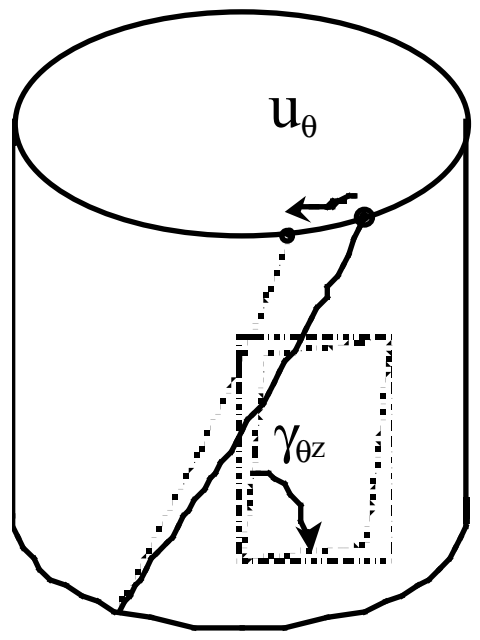

**Figure 6.** *Torsion de la cellule unique et isolée dans le cas d'un angle des microfibrilles*

Cette résolution serait transposable au cas de la fibre unique et isolée. Les déformations induites alors par une déformation potentielle entraîneraient notamment une torsion de la cellule. Cette déformation est incompatible avec la réalité de la fibre dans le bois, où les déformations de torsion sont empêchées. En effet, le schéma d'enroulement des microfibrilles est le même dans toutes les fibres , les parois de deux cellules adjacentes sont donc antisymétriques par rapport à la lamelle moyenne, d'un côté l'angle des microfibrilles avec l'axe de la fibre est positif et de l'autre négatif (Figure 7). Cette symétrie entraîne un déplacement tangentiel nul sur la lamelle moyenne incompatible avec une torsion de la fibre. D'autre part une torsion constante de la fibre correspond à un déplacement transversal qui croît avec z, ce qui n'est pas le cas pour la fibre dans le bois. Cette résolution ne rendrait donc pas compte du comportement de la fibre dans le bois.

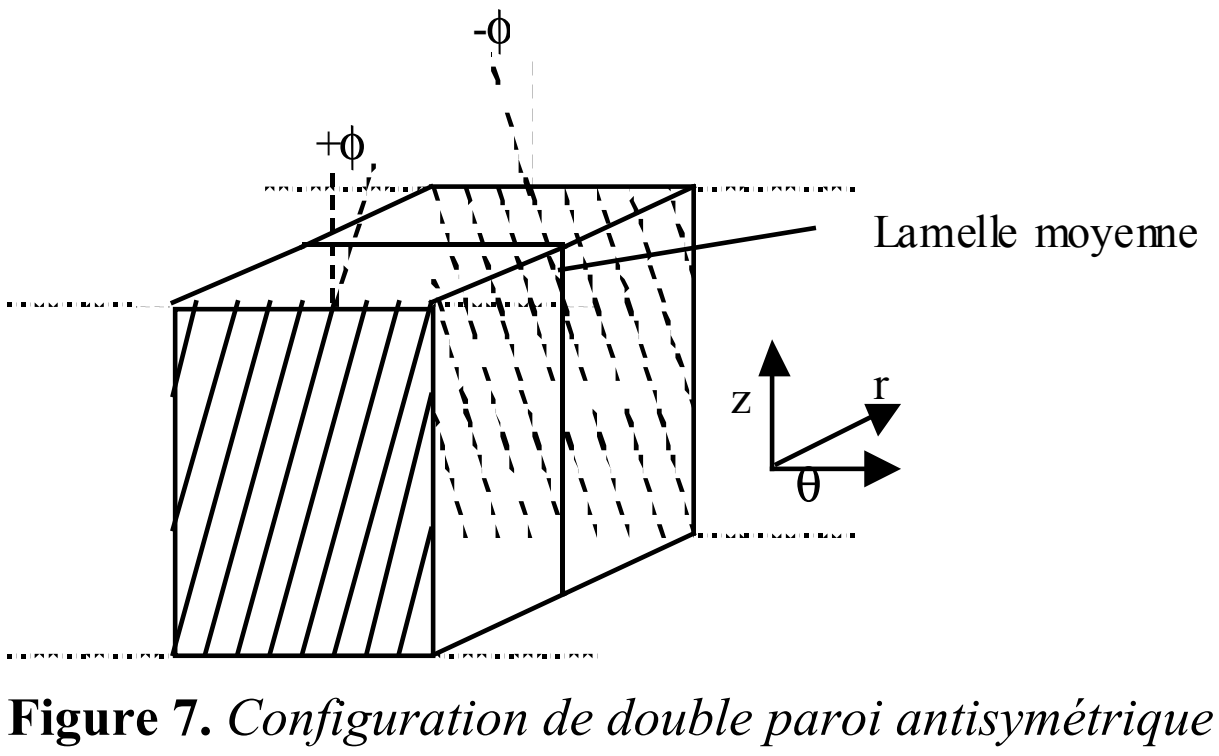

**Figure 7.** *Configuration de double paroi antisymétrique*

Voyons maintenant la situation de **la cellule in-situ**. Les rigidités et les déformations dans le repère {r,θ,z} (équation 2) peuvent être décomposées en partie paire (exposant p) et impaire (exposant *) par rapport à l'angle des microfibrilles. Les composantes orthotrope et anisotrope des rigidités s'écrivent dans le repère local de la fibre :

$$\underline{\underline{C}} = \begin{vmatrix} C_{rr} & C_{r\theta} & C_{rz} & C_{r\rho} & & \\ C_{\theta r} & C_{\theta\theta} & C_{\theta z} & C_{\theta\rho} & & \\ C_{zr} & C_{z\theta} & C_{zz} & C_{z\rho} & & \\ C_{\rho r} & C_{\rho\theta} & C_{\rho r} & C_{\rho\rho} & & \\ & & & & C_{\tau\tau} & C_{\tau\lambda} \\ & & & & C_{\lambda\tau} & C_{\lambda\lambda} \end{vmatrix} = \underline{\underline{C}}^p + \underline{\underline{C}}^*$$

$$\begin{vmatrix} C_{rr} & C_{r\theta} & C_{rz} & & & \\ C_{\theta r} & C_{\theta\theta} & C_{\theta z} & & & \\ C_{zr} & C_{z\theta} & C_{zz} & & & \\ & & & C_{\rho\rho} & & \\ & & & & C_{\tau\tau} & \\ & & & & & C_{\lambda\lambda} \end{vmatrix} \begin{vmatrix} & & & C_{r\rho} & & \\ & & & C_{\theta\rho} & & \\ & & & C_{z\rho} & & \\ C_{\rho r} & C_{\rho\theta} & C_{\rho r} & & & \\ & & & & & C_{\tau\lambda} \\ & & & & C_{\lambda\tau} & \end{vmatrix}$$

La décomposition des déformations potentielles en valeurs normales $\alpha_n$ et torsion $\alpha_{\theta z}$ est sous forme vectorielle :

$$\underline{\alpha} = \underbrace{\begin{vmatrix} \alpha_{rr} \\ \alpha_{\theta\theta} \\ \alpha_{zz} \\ \alpha_{\theta z} \\ 0 \\ 0 \end{vmatrix}}_{\alpha} \underbrace{= \underline{\alpha}^p + \underline{\alpha}^* =}_{\alpha^n} \begin{vmatrix} \alpha_r \\ \alpha_\theta \\ \alpha_z \\ 0 \\ 0 \\ 0 \end{vmatrix} \underbrace{+}_{\underline{\alpha}^c} \begin{vmatrix} 0 \\ 0 \\ 0 \\ \alpha_\rho \\ 0 \\ 0 \end{vmatrix}$$

La déformation de cisaillement de la fibre dans le bois $\gamma_{\theta z}$ est nulle en périphérie. Dans le cadre considéré des déformations planes, ce cisaillement est nul dans toute la fibre. Le comportement anisotrope de la paroi est dû à l'orientation des microfibrilles dans le plan local $\theta z$. Il introduit un couplage entre les déformations de cisaillement dans ce plan et les contraintes normales $\underline{\sigma}^p=(\sigma_r, \sigma_\theta, \sigma_z, 0, 0, 0)$. La loi de comportement élastique avec déformation induite peut être réduite aux contraintes normales exprimées en fonction des déformations normales $\underline{\varepsilon}^p=(\varepsilon_r, \varepsilon_\theta, \varepsilon_z, 0, 0, 0)$ et de cisaillement $\underline{\varepsilon}^*=(0, 0, 0, \gamma_{\theta z}, 0, 0)$ dans le plan $\theta z$ :

$$\sigma^p = \underline{\underline{C}}^p(\varepsilon^p-\alpha^p) + \underline{\underline{C}}^*(\varepsilon^*-\alpha^*)$$

$C^p$ est la partie orthotrope cylindrique du comportement et $\underline{\underline{C}}^*$ est la partie représentant le couplage entre les composantes normales et les cisaillements. La première est un terme pair en fonction de l'angle des microfibrilles, alors que la seconde est un terme impair. Cette loi de comportement, compte-tenu de la restriction de cisaillement imposant $\gamma_{\theta z}=0$, peut être rapportée pour les contraintes et déformations normales à une loi de comportement orthotrope cylindrique en écrivant :

$$\sigma^p = \underline{\underline{C}}^p(\varepsilon^p-\alpha^э) \text{ avec } \alpha^э= \alpha^p+ (\underline{\underline{C}}^p)^{-1}\underline{\underline{C}}^*\alpha^* \qquad [2]$$

Dans la suite, pour alléger les notations, nous ne considérerons que les parties normales et omettrons l'exposant p : $\sigma$, $\varepsilon$, $\underline{\underline{C}}$ et $\alpha$ désigneront $\sigma^p$, $\varepsilon^p$, $\underline{\underline{C}}^p$ et $\alpha^э$, respectivement. Les contraintes ainsi calculées ne seront pas les contraintes réelles dans les couches, pour les connaître il conviendrait d'ajouter à chaque fois le terme complémentaire : $\sigma^* = \underline{\underline{C}}^*\varepsilon + \underline{\underline{C}}^p\alpha^* + \underline{\underline{C}}^*\alpha^p$

*2.3.3. Formulation du problème mécanique*

Compte tenu des hypothèses citées plus haut, la solution du problème en déplacement est cherchée sous la forme $u_r=u(r)$ $u_\theta=0$ et $u_z=\Delta\varepsilon_z$ où $\Delta\varepsilon_z$ est une constante. Les composantes de cisaillement sont nulles. Les équations du problème mécanique s'écrivent :

**Equilibre statique :** $\text{div}(\underline{\sigma})=0 \rightarrow \sigma_r' + \dfrac{\sigma_r - \sigma_\theta}{r} = 0$

**Compatibilité cinématique :**

$$\underline{\varepsilon}=1/2(\text{grad}(\underline{u})+{}^t\text{grad}(\underline{u})) \rightarrow \varepsilon_\theta=u_r/r \text{ et } \varepsilon_r=u_r'$$

**Comportement** : $\underline{\sigma}=\underline{\underline{C}}^i(\underline{\varepsilon}-\underline{\alpha}^i)$

où les $\underline{\underline{C}}^i$ et $\underline{\alpha}^i$ sont données pour chaque couche par les équations [1] et [2]

**Conditions aux limites.** La fibre est libre de contrainte à l'intérieur du lumen ($r=r_n$). Les contraintes à la périphérie de la fibre ($r=r_0$) pourraient être estimées à partir des contraintes extérieures macroscopiques exercées sur le volume de bois. On considère ces dernières nulles dans le cas du retrait et de la maturation libres de contraintes extérieures ; la contrainte radiale et longitudinale (équation 3) sur la fibre rendant compte des déformations du bois est donc supposée nulle aussi :

$$\sigma_r(r_0) = \sigma_r(r_n) = 0 \text{ et } \int \sigma_z ds = 0 \qquad [3]$$

**Interfaces et périphérie.** Il n'y a pas de décollement entre les couches (Figure 8). Cette condition entraîne que le déplacement radial est continu à l'interface $r=r^i$ :

$$u^i_r(r^i) = u^{i+1}_r(r^i) \text{ pour } i=1,..,n-1 \qquad [4]$$

De même, la contrainte radiale est continue pour chaque interface :

$$\sigma^i_r(r^i) = \sigma^{i+1}_r(r^i) \text{ pour } i=1,..,n-1 \qquad [5]$$

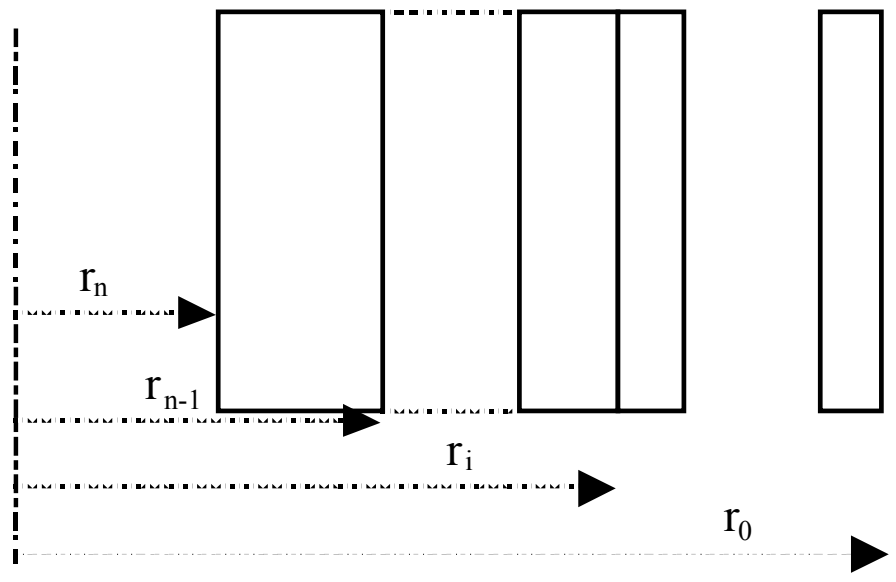

**Figure 8.** *Conditions limites à la périphérie et aux interfaces*

### *2.4 Résolution*

Archer [ARC 86] propose une résolution analytique de la déformation induite dans un matériau orthotrope cylindrique dans le cadre des déformations planes généralisées et dans le cas où il n'y aurait pas de cisaillement dans le plan θz.

Cette résolution est appliquée à chaque couche.

#### *2.4.1. Solution analytique dans chaque couche de la paroi*

L'équation d'équilibre revient à chercher une fonction contrainte $\phi$ telle que $\sigma_r=\phi/r$ et $\sigma_\theta = \phi'$. Compte tenu de la compatibilité cinématique et de la loi de comportement, cette fonction vérifie alors l'équation différentielle linéaire du second ordre :

$$r\phi''+\phi'-\lambda^2\frac{\phi}{r}=P+Q\Delta\varepsilon_z \text{ avec } P=\frac{\alpha_r'-\alpha_{\theta r}'}{\beta_{22}},\ Q=\frac{S_{rz}-S_{\theta z}}{\beta_{22}S_{zz}} \text{ en notant}$$

$$\beta_{11}=S_{rr}-S_{rz}^2/S_{zz}\ ;\ \beta_{12}=S_{r\theta}-S_{rz}S_{\theta z}/S_{zz}\ ;\ \beta_{22}=S_{\theta\theta}-S_{\theta z}^2/S_{zz}$$

$$\alpha_r'=\alpha_r-S_{rz}\alpha_z/S_{zz}\ ;\ \alpha_\theta'=\alpha_\theta-S_{\theta z}\alpha_z/S_{zz}$$

$$\lambda=\beta_{11}/\beta_{22}$$

Deux cas sont envisagés dans la résolution :

$\lambda\neq 1$ : $\phi$ est de la forme $\phi = A\, r^{\lambda} + B\, r^{-\lambda} + (P+Q\Delta\varepsilon_z)\, r / (1-\lambda^2)$

$\lambda=1$ : $\phi$ est de la forme $\phi = A\, r^{\lambda} + B\, r^{-\lambda} + (P+Q\Delta\varepsilon_z)\, r \ln(r) / 2$

A et B sont deux constantes inconnues. $\Delta\varepsilon_z$ est le déplacement suivant z constant dans la couche et dans la fibre.

*2.4.2. Résolution numérique pour le multicouche*

Dans chaque couche i ($1 \le i \le n$), le potentiel est donné par les deux constantes $A^i$ et $B^i$ et par la constante commune à toutes les couches $\Delta\varepsilon_z$. Il permet alors de calculer la contrainte radiale $\sigma_r^i$ et transversale $\sigma_\theta^i$ et donc le déplacement radial :

$$u_r^i = r\,\varepsilon_\theta = r\,(\beta_{12}^i\,\sigma_r^i + \beta_{22}^i\,\sigma_\theta^i + \alpha_\theta'^i + S_{23}^i/S_{33}^i\,\Delta\varepsilon_z).$$

La contrainte radiale et le déplacement radial s'expriment linéairement en fonction des constantes $A^i$, $B^i$ et $\Delta\varepsilon_z$. Le système linéaire ainsi formé comprend 2n+1 inconnues. Les interfaces entre les couches définissent n-1 conditions aux limites pour le déplacement radial (équation 4) et n-1 conditions aux limites pour la contrainte radiale (équation 5). A ces 2n-2 équations, s'ajoutent les conditions de surface externe libre (équation 3). Le système linéaire est équilibré ; il comprend 2n+1 équations et inconnues. Les équations sont écrites sous la forme :

Continuité de la contrainte radiale à l'interface $r_i$ ($1\le i\le n-1$) :

$$r_i^{\lambda_i-1}A_i + r_i^{-\lambda_i-1}B_i + k_i(r_i)\Delta\varepsilon_z + m_i(r_i) = r_i^{\lambda_{i+1}-1}A_{i+1} + r_i^{-\lambda_{i+1}-1}B_{i+1} + k_{i+1}(r_i)\Delta\varepsilon_z + m_{i+1}(r_i)$$

Continuité du déplacement radial à l'interface $r_i$ ($1\le i\le n-1$) :

$$f_i(r_i)A_i + g_i(r_i)B_i + l_i(r_i)\Delta\varepsilon_z + p_i(r_i) = f_{i+1}(r_i)A_{i+1} + g_{i+1}(r_i)B_{i+1} + l_{i+1}(r_i)\Delta\varepsilon_z + p_{i+1}(r_i)$$

La nullité de la résultante suivant z s'écrit sous la forme :

$$\sum_{i=1,n}(j_i(r_i)-j_i(r_{i-1}))A_i+(h_i(r_i)-h_i(r_{i-1}))B_i$$
$$+\left(\sum_{i=1,n}(q_i(r_i)-q_i(r_{i-1}))\right)\Delta\varepsilon_z+\left(\sum_{i=1,n}(s_i(r_i)-s_i(r_{i-1}))\right)=0$$

Ce système linéaire est construit à partir des fonctions définies comme suit :

| | $\lambda\neq 1$ | $\lambda=1$ |
|---|---|---|
| | $(\beta_{12}+\lambda\beta_{22})\, r^{\lambda-1}$ | |
| | $(\beta_{12}-\lambda\beta_{22})\, r^{-\lambda-1}$ | |
| | $[2(S_{rz}-S_{\theta z})\frac{1}{\lambda+1}+2S_{\theta z}]\frac{r^{\lambda+1}}{S_{zz}}$ | |
| | $[2(S_{rz}-S_{\theta z})\frac{1}{-\lambda+1}+2S_{\theta z}]\frac{r^{-\lambda+1}}{S_{zz}}$ | $[2(S_{rz}-S_{\theta z})\ln(r)+2S_{\theta z}]/S_{zz}$ |
| | $Q/(1-\lambda^2)$ | $Q\ln(r)/2$ |
| | $Q(\beta_{12}+\beta_{22})/(1-\lambda^2)+S_{\theta z}/S_{zz}$ | $Q(\beta_{12}\ln(r)+\beta_{22}(\ln(r)+1))/2+S_{\theta z}/S_{zz}$ |
| | $P/(1-\lambda^2)$ | $P\ln(r)/2$ |
| | $P(\beta_{12}+\beta_{22})/(1-\lambda^2)+\alpha_\theta'$ | $P(\beta_{12}\ln(r)+\beta_{22}(\ln(r)+1))/2+\alpha_\theta'$ |
| | $[Q(S_{rz}+S_{\theta z})/(1-\lambda^2)-1]\, r^2/S_{zz}$ | $[Q(S_{rz}+S_{\theta z})\ln(r)/2-Q(S_{rz}-S_{\theta z})/4-1]\, r^2/S_{zz}$ |
| | $[P(S_{rz}+S_{\theta z})/(1-\lambda^2)+\alpha_z]\, r^2/S_{zz}$ | $[P(S_{rz}+S_{\theta z})\ln(r)/2-Q(S_{rz}-S_{\theta z})/4+\alpha_z]\, r^2/S_{zz}$ |

Le système linéaire est alors écrit sous la forme MX+P=0 (Figure 10). X est le vecteur des 2n+1 inconnues, P le vecteur des seconds membres et M la matrice carrée de dimension 2n+1. Ces deux derniers dépendent de la géométrie ($r_i$), des caractéristiques mécaniques et des potentiels de retrait.

Les inconnues $A_i$, $B_i$ et $\Delta\varepsilon_z$ sont calculées numériquement par résolution du problème linéaire : $X=-M^{-1}P$. On obtient ainsi la déformation longitudinale de la fibre $\varepsilon_L=\Delta\varepsilon_z$ et la déformation transversale :

$$\varepsilon_T=u_r(r_n)/r_n=\varepsilon_\theta(r_n).$$

Les équations sur l'ensemble des couches peuvent se ramener à l'expression matricielle indiquée sur la Figure 9.

*2.4.3. Cinétiques de déformation induite et de rigidification*

La cinétique est la conséquence du caractère progressif de la lignification au sein de la paroi. Ce processus est illustré sur la Figure 9 est supposé :

- Centripète : la lignification commence par la lamelle moyenne et la paroi primaire. Une fois achevée dans cette paroi elle se poursuit dans S1, ainsi de suite…

- Responsable des déformations induites dans la matrice mais aussi dans le réseau microfibrillaire ; ces déformations sont décrites par des courbes sigmoïdes. La lignification entraîne le gonflement de la matrice [BOY 72]. On suppose aussi qu'elle amorce la cristallisation de la cellulose et donc le retrait du réseau microfibrillaire [BAM 78].

- Augmenter la rigidité de la matrice : le module d'Young est quasiment nul au début de la lignification et croît pour atteindre sa valeur finale Em en fin de lignification.

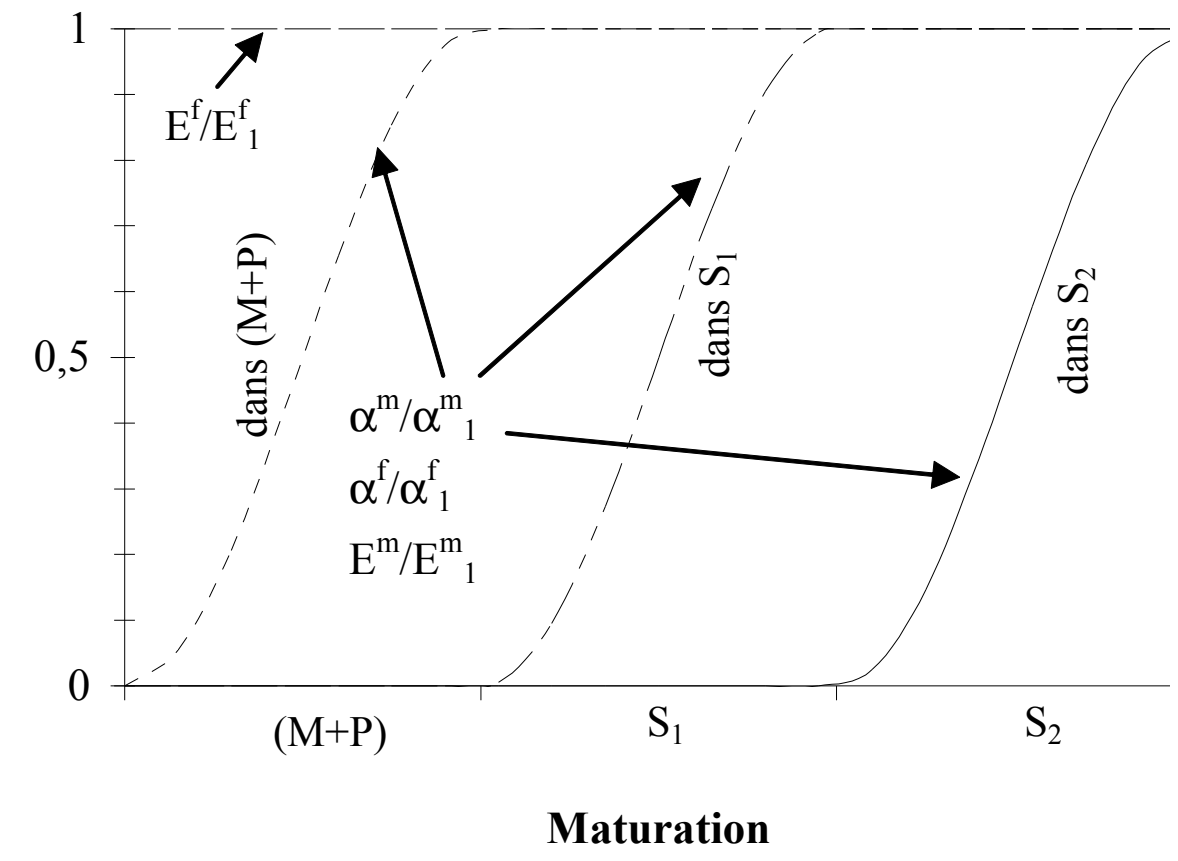

***Figure 9.*** *Cinétique des déformations induites et des caractéristiques mécaniques pendant la maturation*

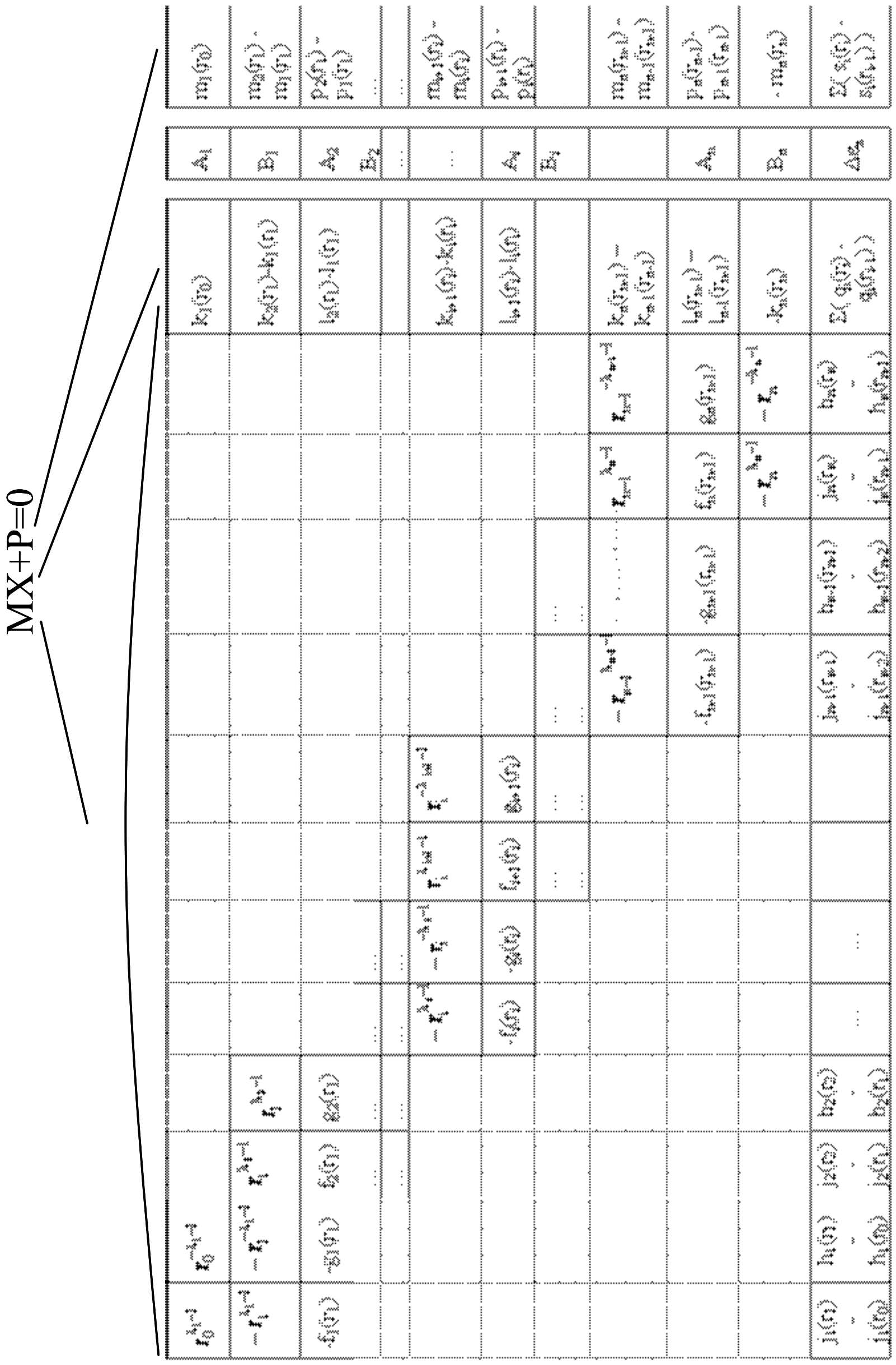

***Figure 10.*** *Obtention des constantes du problème par résolution d'un système linéaire*

### *2.5 Remarques sur le modèle*

Le modèle mécanique ainsi construit permet de déterminer les déformations de la cellule en fonction de sa structure, c'est à dire des paramètres que sont la géométrie, la composition des différentes couches et les caractéristiques des constituants de la matière ligneuse. Ces derniers peuvent être supposés structurels, c'est à dire qu'ils ne varient pas d'une couche à l'autre ou d'une nature de fibre à une autre. Ce sont des constantes qui caractérisent la structure et la composition chimique de la matrice et du réseau microfibrillaire. Les différences de déformations observées entre les différents types de cellules devront alors être expliquées par des différences de géométrie ou de composition.

- les rayons des différentes couches et l'angle des microfibrilles dans la paroi $S_2$ peuvent être mesurés par microscopie,
- les différents constituants peuvent être dosés, mais uniquement en moyenne sur la cellule.

Certaines mesures des propriétés mécaniques de la lignine et de la cellulose ont été réalisées, ils dépendent de l'état considéré (température et humidité) mais aussi de la méthode d'extraction. Les constituants considérés dans le modèle sont des organisations en réseau de ces macromolécules. Les caractéristiques mécaniques du réseau peuvent être différentes de celles des macromolécules.

Les potentiels de déformations de la matrice et du réseau microfibrillaire pendant la maturation sont difficilement estimables. Ils deviendront les paramètres d'ajustement du modèle entre les données expérimentales et la simulation. Les trachéides de résineux ont fait l'objet de nombreuses mesures. La modélisation des déformations de maturation de ces cellules doit permettre de caler ces paramètres.

## 3. Modélisation des déformations longitudinales de maturation

### *3.1 Géométrie structure et composition chimique pour la fibre standard*

Dans le cas de fibres ne comportant pas de couche G, la paroi $S_3$ de faible épaisseur est négligée, de sorte qu'on peut se contenter d'un modèle à 3 couches. Les rayons des couches et les proportions des constituants sont pris identiques dans toutes les simulations. L'angle des microfibrilles dans les couches (M+P) et $S_1$ est constant. Il varie par contre dans $S_2$ de 0° à 60°. L'angle de 20° représente la fibre de bois normal. L'angle de 60° correspond au cas extrême de la fibre de bois de réaction des résineux. Le cas limite de l'angle nul représente lui la fibre de bois de tension de feuillu ne différenciant pas de couche G. Les caractéristiques mécaniques utilisées par la suite sont celles qui permettent de rendre compte des rigidités de la paroi, tout en assurant un module transverse du réseau microfibrillaire faible devant celui de la matrice (***Tableau 3***). Les rayons des différentes couches sont pris de façon à vérifier : (i) une proportion volumique des couches respectives de 20% pour (M+P), 10% pour $S_1$ et 70% pour $S_2$ ; (ii) une densité volumique de 0,4 g/cm$^3$ calculée à partir de la densité de la matière ligneuse de 1,5 g/cm$^3$. Les proportions des constituants dans chaque couche sont inspirées des modèles cités précédemment [YAM 95, HAR 98]. Elles correspondent à une proportion moyenne de réseau microfibrillaire de 35%. Cette proportion représente de façon réaliste le pourcentage de cellulose cristalline. L'ensemble des paramètres est reporté dans le Tableau 5. Des variations concomitantes des rayons ou des compositions avec les autres caractéristiques ne sont pas représentées dans le modèle, même si l'on sait par exemple que le bois de compression a une déformation longitudinale de maturation forte, un angle des microfibrilles élevé dans $S_2$ et que ses parois sont plus épaisses et plus lignifiées que celle des trachéides de bois normal.

**Tableau 5.** *Paramètres du modèle à 3 couches*

| | S2 | S1 | M+P |
|---|---|---|---|
| Rayon réduit $r_i/r_0$ | 0,86 | 0,96 | 0,98 |
| Proportion du réseau microfibrillaire | 0,45 | 0,2 | 0,1 |

### *3.2 Un modèle à 4 couches pour la fibre G*

Dans le cas de bois de tension à fibres G, il peut être nécessaire d'ajouter une quatrième couche. Les couches M+P et S1 sont alors reprises du modèle précédent. Dans le cas du Hêtre ou du Peuplier, la couche gélatineuse remplace la couche S3 [WIC 79]. Pour le Peuplier, Okumara et al. [OKU 77] citent une épaisseur de 0,2 µm pour la couche S1, 0,4 µm pour S2 et 1,2µm pour la couche G. Pour cette essence, Ritter et al. [RIT 93] estiment à 80% la surface occupée par la couche G. Des épaisseurs de parois sont citées dans la littérature pour le Wapa (*Epurea falcata* L.) en fonction de la nature de la fibre [MAR 83, DEL 94]. Cette dernière compare les épaisseurs de la couche S2 et G et constate une diminution de l'épaisseur de la couche S2 dans le bois de tension.

Des observations au microscope électronique à balayage confirment cette tendance (Figure 11). Les fibres de bois prélevées sous une faible valeur de la déformation résiduelle ont en majorité une structure de fibre de bois normal. Il est difficile de distinguer toutes les couches sur de telles images. La séparation entre la paroi primaire et la paroi secondaire est toutefois nette et permet de mesurer M+P (0,5 µm) et $S_1$+$S_2$ (4 µm). A noter dans le bois de forte déformation résiduelle, une fibre gélatineuse caractéristique par son lumen de faible épaisseur en forme “d'étoile”. La séparation entre la paroi S2 et la couche G est visible dans ce cas. Les deux couches sont d'épaisseur semblable, 3,6 µm pour la couche $S_2$ et 3,2 µm pour la couche G. Sur le bois prélevé sous une zone de très forte déformation résiduelle, les couches gélatineuses apparaissent en nombre, le lumen est quasiment inexistant. Il faut noter que ces fibres résistent mal à l'exposition des électrons et semblent dégradées par l'observation. La couche S2 est mince, elle mesure entre 1 et 2 µm. La couche G peut atteindre 4 µm.

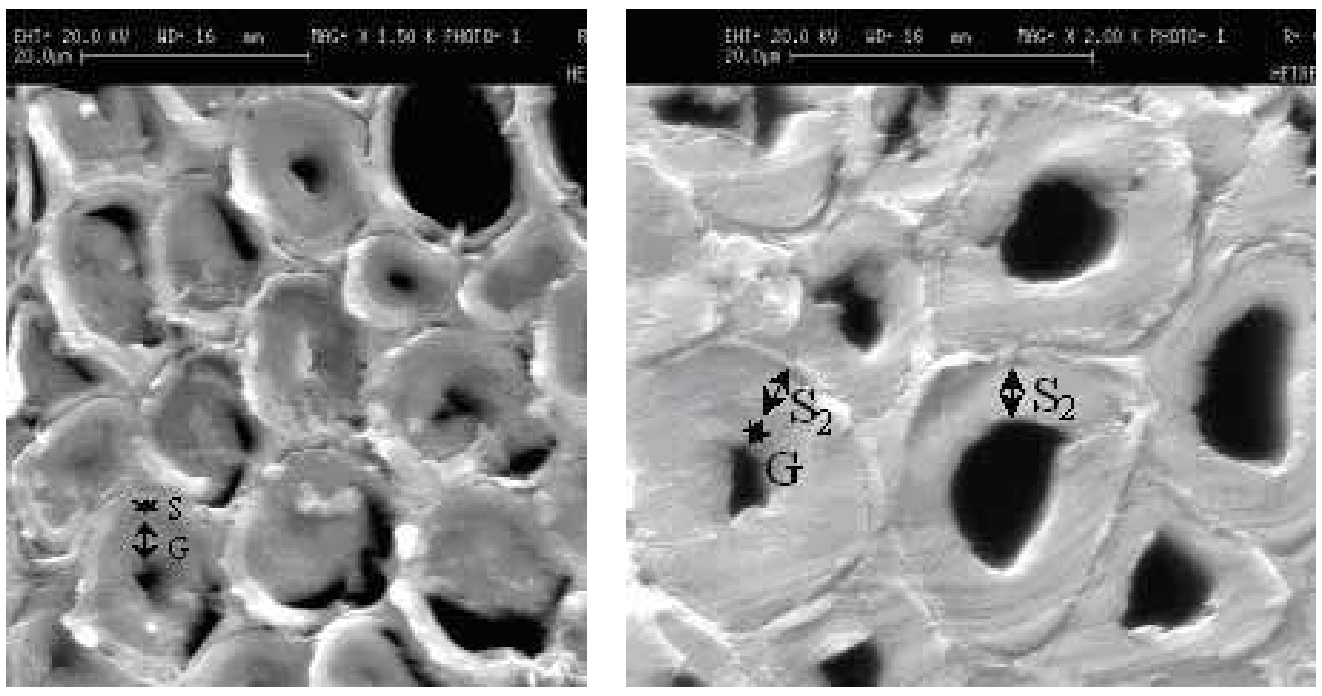

**Figure 11.** *Observation au microscope électronique de bois de Hêtre de faible (gauche) et de forte (droite) déformation résiduelle longitudinale*

Les variations géométriques de la fibre sont alors décrites :

- pour le bois normal, par une augmentation de l'épaisseur de la couche $S_2$ de 2 µm à 4 µm,
- pour le bois de tension, par une augmentation de l'épaisseur de la couche G de 0 µm à 4 µm, et conjointement une diminution de l'épaisseur de la couche $S_2$ de 4 µm à 1 µm.

Pour les deux natures de fibre, le rayon total de la fibre est de 8 µm, le rayon interne de la couche M+P est 7,8 µm , celui de $S_1$ 7,1 µm. L'angle des microfibrilles dans la paroi S2 du BT est peu mesuré. Chez le Hêtre, un angle d'environ 20° est observé par Trénard [TRE 81].

### *3.2 Modélisation de la déformation de maturation*

Le modèle à trois parois est utilisé pour calculer les déformations de maturation des trachéides de résineux et des fibres de bois de tension des feuillus ne différenciant pas de couche G.

La déformation potentielle de la matrice correspond au gonflement isotrope de la lignine alors que la déformation potentielle du réseau microfibrillaire correspond au retrait isotrope transverse des microfibrilles de cellulose.

Les données expérimentales regroupent :

- les déformations longitudinales de maturation mesurées en périphérie des arbres sur pied
- l'angle des microfibrilles dans S2.

Elles ont été mesurées sur trois espèces de résineux, un feuillu ne différenciant pas de couche gélatineuse dans son bois de tension (eucalyptus) et sur du buis qui, par ses déformations longitudinales de maturation et sa composition chimique, s'apparente à un résineux. Les tendances sont schématisées par trois nuages (Figure 12) :

- Bois de compression : angle de microfibrille supérieur à 25° et déformation longitudinale de maturation en gonflement.
- Bois Normal : angle de microfibrille dans la plage [10°,25°] et retrait longitudinal de maturation compris entre 0 et –0,1%.
- Bois de Tension : angle de microfibrille inférieur à 10° et déformation longitudinale de maturation inférieure (en valeur absolue) à –1%.

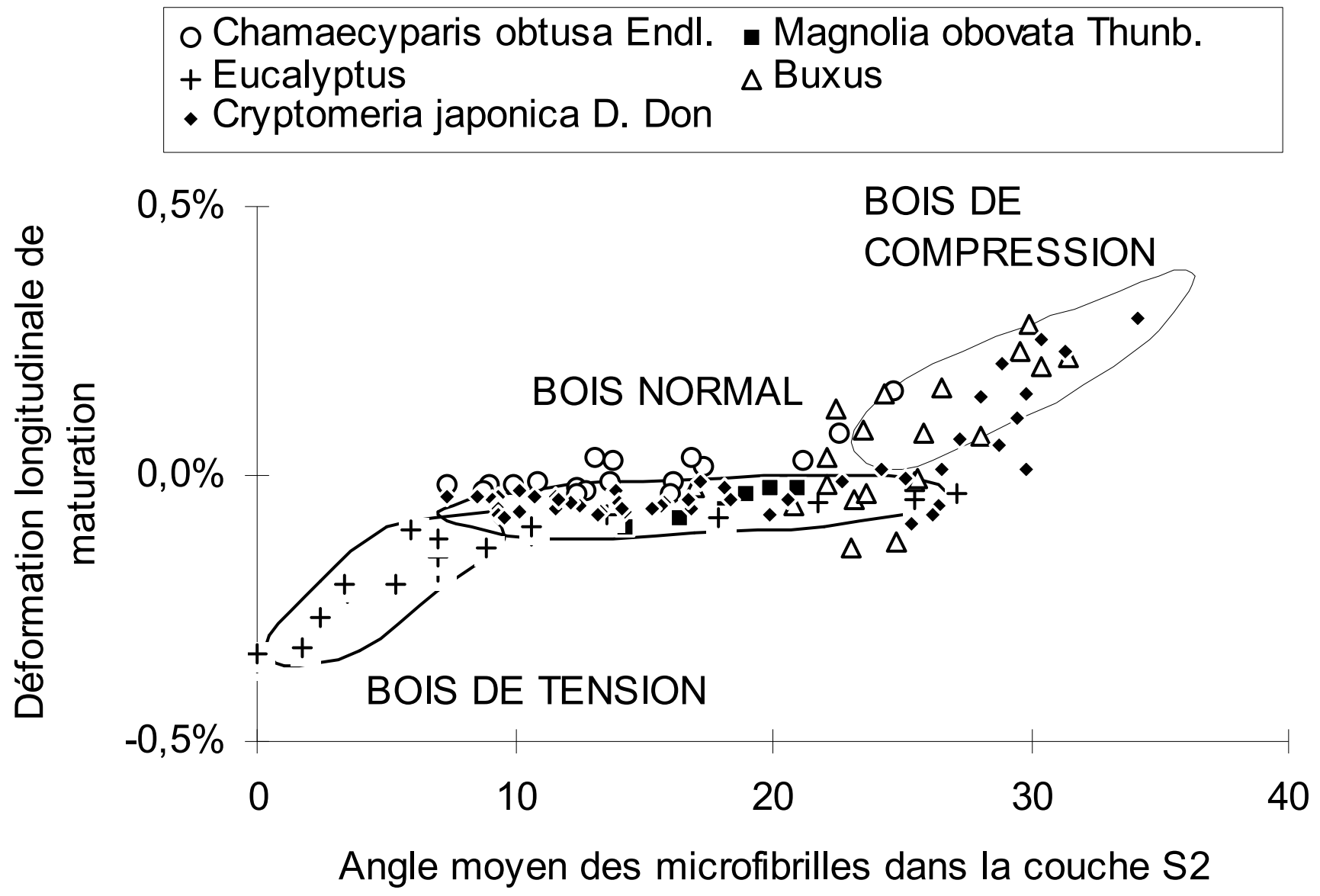

**Figure 12.** *Relation entre angle des microfibrilles dans la couche S2 et déformation longitudinale de maturation.*

Pour illustrer ces tendances, nous décrivons l'ensemble des variations de la structure de la fibre entre le BN et le BT. Les variations de la composition chimique sont prises en compte par un changement de taux de cellulose dans la paroi $S_2$. Ce paramètre est ajusté de telle façon que le pourcentage de réseau microfibrillaire sur l'ensemble de la cellule varie à peu près continûment de 30% pour le bois de compression à 45% pour le bois de tension. Ces variations ne sont pas exactement celles de la cellulose, puisque par exemple la proportion lignine + hemicellulose peut atteindre 55% pour le bois de compression de *pinus pinaster* L. contre 45% pour le bois normal [COM 97] alors que pour l'eucalyptus, le taux de lignine de 32% pour le bois normal chute de 10% dans le bois de tension [BAI 94]. Pour ne pas compliquer la description des écarts de composition du bois normal des différentes essences, nous avons représenté uniquement la tendance commune qui est une augmentation du taux de cellulose du bois de compression au bois de tension.

Plusieurs sous-groupes sont considérés pour chaque nature de cellule (bois de compression, bois normal ou bois de tension) pour lesquels sont définis des paramètres différents (Tableau 6).

**Tableau 6.** *Paramètres de la modélisation de la relation entre l'AMF et la déformation résiduelle longitudinale*

| | $\alpha^m$ | $\alpha^f_1$ | % réseau dans $S_2$ |
|---|---|---|---|
| BC ++ | 1,50% | -0,5‰ | 0,35 |
| BC + | 1,25% | -0,75‰ | 0,40 |
| BC | 1% | -1‰ | 0,45 |
| BN + | 0,75% | -1,25‰ | 0,50 |
| BN - | 0,50% | -1,5‰ | 0,50 |
| BT | 0,50% | -2‰ | 0,55 |
| BT - | 0,25% | -3‰ | 0,55 |
| BT -- | 0,25% | -4‰ | 0,60 |

Les déformations de maturation du bois de tension de feuillus différenciant une couche gélatineuse sont modélisées en introduisant une quatrième couche. La description des variations de la géométrie et de la composition chimique du bois permet de modéliser la déformation longitudinale de maturation et donc la relation entre ces deux propriétés observées expérimentalement pour le Hêtre et le Peuplier. Pour le Hêtre et le Peuplier, le retrait de maturation du bois de tension est supérieur à celui du bois normal. Pour rendre compte de cette différence, les déformations potentielles dans la couche G sont prises supérieures aux valeurs des autres parois. Il est difficile d'observer, même qualitativement les différences de comportement de la couche gélatineuse. Par contre les déformations potentielles de la couche G sont prises différentes. Cette caractéristique particulière vise à rendre compte du comportement différent de la fibre de bois de tension, comparativement à la fibre de bois normal. Les déformations potentielles de maturation dans les couches M+P, S1 et S2 sont celles utilisées pour le bois normal dans le paragraphe précédent (ligne BN- au Tableau 6).

La simulation de la relation entre l'AMF et la déformation longitudinale de maturation de la fibre nécessite de faire varier les paramètres de déformations potentielles considérés auparavant comme structuraux. L'ajustement est obtenu par une diminution du gonflement de la matrice du bois de compression au bois de tension conjointement à une augmentation du retrait du réseau microfibrillaire (Figure 13).

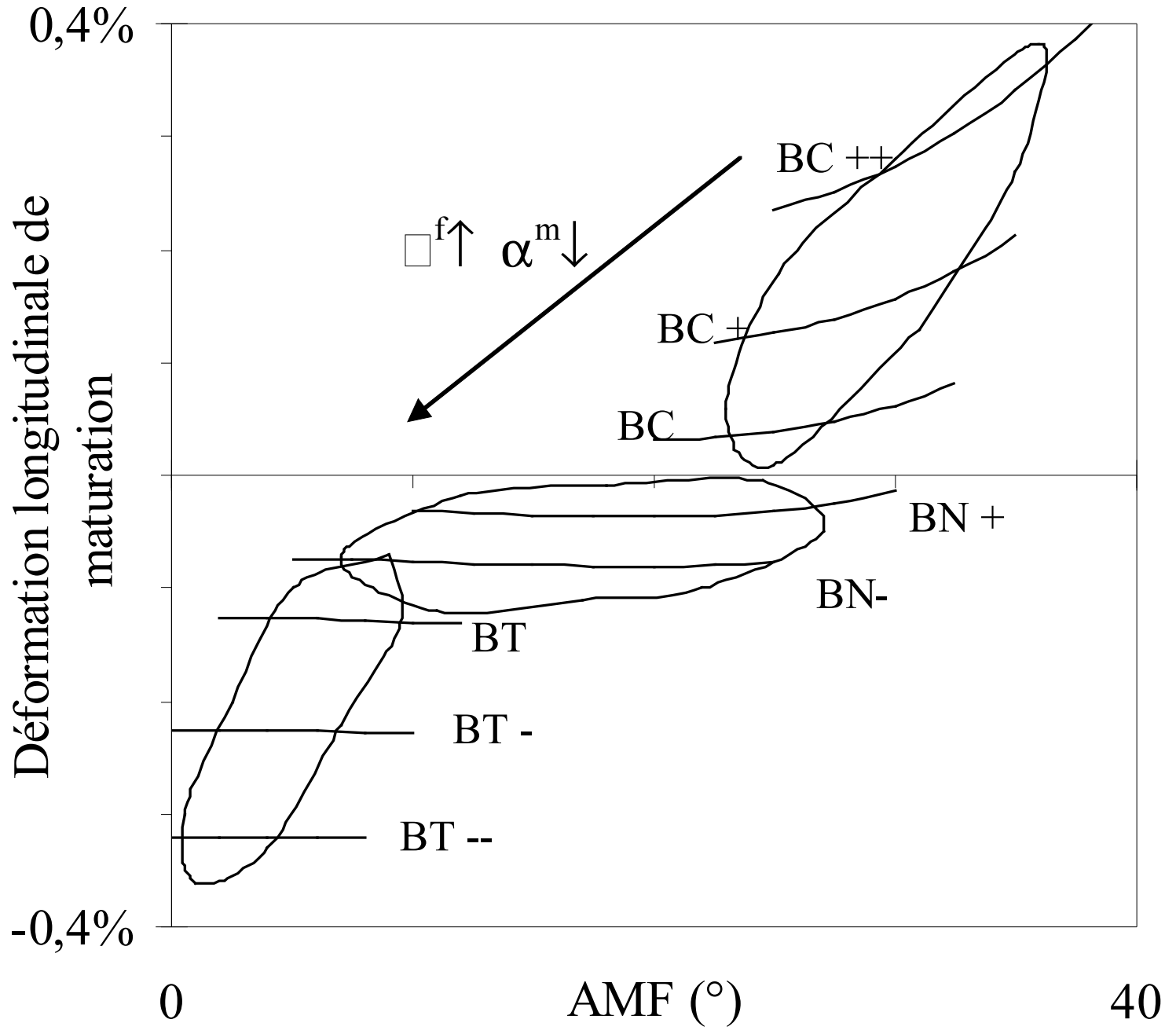

**Figure 13**. *Modélisation de la relation entre l'AMF et la déformation résiduelle longitudinale*

**Conclusion**

Nos simulations suggèrent que les variations de l'angle des microfibrilles seules ne peremttent pas, avec le modèle retenu, de rendre compte des variations de la déformation longitudinale. Il est indispensable de supposer également des variations de déformations induites par la maturation, correspondant à des différences de composition chimique éventuellement corrélées à l'angle des microfibrilles. Nous avons supposé dans ce calcul que les variations de rigidité de la matrice se produisent de manière parfaitement synchrone avec l'apparition des déformations induites. Cette hypothèse pourrait raisonnablement être remise en cause, ce qui pourrait aboutir à des résultats et conclusions assez différents.

## Bibliographie